\documentclass[12pt]{article}
\usepackage{enumitem}
\usepackage{fullpage}
\begin{document}
\title{A Lower Bound for Boolean Satisfiability\\on Turing Machines}
\author{Samuel C. Hsieh\\Computer Science Department, Ball State University}
\maketitle
\begin{abstract}
We establish a lower bound for deciding the satisfiability of the conjunction of any two Boolean formulas from a set called a \textit{full representation} of Boolean functions of $n$ variables - a set containing a Boolean formula to represent each Boolean function of $n$ variables.  The contradiction proof first assumes that there exists a Turing machine with $k$ symbols in its tape alphabet that correctly decides the satisfiability of the conjunction of any two Boolean formulas from such a set by making fewer than $2^nlog_k2$ moves. By using multiple runs of this Turing machine, with one run for each Boolean function of $n$ variables, the proof derives a contradiction by showing that this Turing machine is unable to correctly decide the satisfiability of the conjunction of at least one pair of Boolean formulas from a  full representation of $n$-variable Boolean functions if the machine makes fewer than $2^nlog_k2$ moves. This lower bound holds for any full representation of Boolean functions of $n$ variables, even if a full representation consists solely of minimized Boolean formulas derived by a Boolean minimization method. We discuss why the lower bound fails to hold for satisfiability of certain restricted formulas, such as 2CNF satisfiability, XOR-SAT, and HORN-SAT. We also relate the lower bound to 3CNF satisfiability. The lower bound does not depend on sequentiality of access to the tape squares and will hold even if a machine is capable of non-sequential access. 
\end{abstract}
\section{Introduction}
The problem of deciding whether a Boolean formula is satisfiable is commonly known as the Boolean satisfiability problem. It was the first problem shown to be NP-complete [1]. This paper establishes a lower bound for deciding the satisfiability of the conjunction of any two Boolean formulas from a set called a \textit{full representation} of Boolean functions of $n$ variables - a set containing a Boolean formula to represent each Boolean function of $n$ variables.  The contradiction proof first assumes that there exists a Turing machine with $k$ symbols in its tape alphabet that correctly decides the satisfiability of the conjunction of any two Boolean formulas from such a set by making fewer than $2^nlog_k2$ moves. By using multiple runs of this Turing machine, with one run for each Boolean function of $n$ variables, the proof derives a contradiction by showing that this Turing machine is unable to correctly decide the satisfiability of the conjunction of at least one pair of Boolean formulas from a  full representation of $n$-variable Boolean functions if the machine makes fewer than $2^nlog_k2$ moves.

We briefly summarize the remaining sections of this paper. The next section provides a brief overview of Boolean formulas and Turing machines. As there are variations in the nomenclatures used in the literature, this overview of the related concepts and terminology as used in this paper seems appropriate. Section 3  introduces concepts related to  executing multiple runs of a Turing machine and proves a few related lemmas. Section 4 proves the lower bound of $2^nlog_k2$ moves and shows that the lower bound applies to CNF satisfiability and, by duality, DNF falsifiability. Section 5 discusses why the lower bound fails to hold for satisfiability of certain restricted formulas such as 2CNF satisfiability, XOR-SAT and HORN-SAT, and the section then relates the lower bound to 3CNF satisfiability. Section 6 discusses  the lower bound with regard to the number of distinct tape symbols and to non-sequential access to tape squares.
\section{Boolean Formulas and Turing Machines}
Boolean formulas and Turing machines are widely known, e.g., [2,3]. As there are variations in the nomenclatures, we summarize the related concepts and terminology as used here.
\subsection{Boolean Formulas}
The set $B=\{true, false\}$ denotes the set of \textit{Boolean values}. A Boolean \textit{variable} has either $true$ or $false$ as its value. A function $f:B^n$ $\rightarrow$ $B$ is a Boolean function of $n$ variables. The expression $B^n \rightarrow B$ denotes the set of Boolean functions of $n$ variables. 

We may use a Boolean \textit{formula} to define or represent a Boolean function. A Boolean formula is composed of Boolean values, Boolean variables, and the Boolean operators $\wedge$ (for conjunction, i.e., AND), $\vee$ (for disjunction, i.e.,  OR) and $\overline{overbar}$ (for negation, i.e., NOT). A Boolean function can be represented by many different Boolean formulas.

A \textit{literal} is a Boolean variable or a logically negated variable. A Boolean formula in DNF  (\textit{Disjunctive Normal Form}), or a DNF formula, is an OR of DNF clauses, and a DNF clause is an AND of literals. For example, $( \overline{x_1} \wedge x_2 )\vee (x_1 \wedge \overline{x_2}) \vee (x_1 \wedge x_3)$  is a DNF formula. A DNF formula is a $k$DNF formula if each clause has k literals. The example just given is a 2DNF formula. A Boolean formula in CNF (\textit{Conjunctive Normal Form}), or a CNF formula, is an AND of CNF clauses, and a CNF clause is an OR of literals. For example, $(x_1 \vee  x_2 ) \wedge (\overline{x_1} \vee \overline{x_2}) \wedge ( x_2 \vee x_3)$ is a CNF formula. A CNF formula is a $k$CNF formula if each clause has $k$ literals. The example just given is a 2CNF formula.

An \textit{assignment} to a set of Boolean variables assigns a Boolean value to each variable in the set. An assignment can be used to evaluate a Boolean formula or a Boolean function. If an assignment makes a formula or a function $true$,  the assignment is said to \textit{satisfy} the formula or the function and is called a \textit{satisfying} assignment; similarly, if an assignment makes a formula or a function $false$, the assignment is said to \textit{falsify} the formula or the function and is called a \textit{falsifying} assignment.

A Boolean formula is \textit{satisfiable} if it has a satisfying assignment; otherwise, the formula is \textit{unsatisfiable}. A Boolean formula is \textit{falsifiable} if it has a falsifying assignment; otherwise, the formula is \textit{unfalsifiable}.
\subsection{Turing Machines}

We consider only \textit{deterministic Turing machines}, henceforth referred to simply as Turing machines or merely as machines. A Turing machine [4] is a simple model of computation. A Turing machine has access to a two-way tape that extends indefinitely in both directions. The tape is divided into squares. Each square is "capable of bearing a symbol" [4] from a finite \textit{tape alphabet}, which has at least two distinct symbols, including $blank$ as a symbol. The squares are ordinally similar to the series of integers $\cdots$  $-3,$ $-2,$ $-1,$ $0,$ $1,$ $2,$ $3$ $\cdots$. For convenience in our discussion, we will regard the series of integers as the \textit{addresses} of the squares and usually refer to a specific square by its address, e.g., the square at address $x$ or the $x^{th}$ square. A Turning machine uses a read/write head to scan (i.e., read) or to write the tape, one square at a time. The address of the square where the head is positioned is called the \textit{head position}. Besides, at any time, a Turing machine is in one of a finite set of internal \textit{states}. A Turing machine executes a finite program, which is often defined as a \textit{transition function}, denoted by $\delta$ here. The function $\delta$ specifies each step of the finite program by mapping a machine's current state and currently scanned symbol to the operations that the machine is supposed to perform for the step. Specifically, each step is specified in the following way:
\begin{center}
$\delta$(state $q$, symbol $\alpha$) $=$ (new symbol $\beta$, destination state $q_{dest}$, left or right)
\end{center} 
which has the following meaning: if the Turing machine is currently in the state $q$ and its head scans the symbol $\alpha$, then the  machine performs following three actions: 1)write the symbol  $\beta$ to the square at the current head position (thus,  $\beta$ replaces $\alpha$), 2)make a state transition to enter the destination state $q_{dest}$ , and then 3) move the head left or right by one square. Performing these three actions will be referred to as \textit{making a move}. Henceforth, we will call the tuple (state q, symbol $\alpha$) a \textit{state-symbol pair} and the triplet (new symbol, destination state, left or right) a \textit{move}, which specifies the three actions that a machine performs in making a move. In short, the transition function $\delta$ maps a state-symbol pair to a move. The move to be made next by a machine is \textbf{solely} determined by and depends \textbf{only} on the current state-symbol pair. 

A problem to be solved by a Turing machine is represented by a finite number of symbols provided as input on the tape initially (\textbf{before} the machine starts executing its finite program), and those tape squares not used to represent the input are initially blank. For example, for the Boolean satisfiability problem, a Boolean formula may be provided on the tape as a finite string composed from the symbols of an appropriate alphabet. A Turing machine starts executing its finite program from an internal state designated to be the \textit{start state}. Execution of the finite program of a machine  proceeds in the following manner until the machine halts: the start state and the symbol scanned at the initial head position constitute the initial state-symbol pair, which determines the first move, making the first move leads to another state-symbol pair, which determines the second move, making the second move leads to yet another state-symbol pair, which determines the third move, and so forth. 

A machine halts when it enters a \textit{halting state}. There are only two halting states: the \textit{accept state} and the \textit{reject state}, and all other states are \textit{non-halting states}. A Turing machine that halts in the accept state is said to \textbf{accept} its input, and a machine that halts in the reject state is said to \textbf{reject} its input.
\section{Running a Turing Machine on a Bipartite Input}
First, we will define several related terms. 
\begin{flushleft}
\textbf{Definition 1}. A partition is a set of tape squares. 
\end{flushleft}

We will run a Turing machine to decide the satisfiability of the conjunction of two Boolean formulas. Hence, an input will consist of two parts, one for each conjunct. The two parts of an input will be provided in two \textbf{disjoint} partitions. It is trivial to divide the squares of a tape into two disjoint partitions. As an example, one partition may consist of the squares with addresses greater than some arbitrary integer $x$, with the other partition consisting of those squares with addresses $\leq x$. As another example, one partition may consist of those squares with addresses that are even numbers, with the other partition consisting of those with odd addresses.
\begin{flushleft}
\textbf{Definition 2}.  A \textit{bipartite} input consists of two parts that are provided on a tape divided into two disjoint partitions, with each part of the input placed in a separate partition.
\end{flushleft}

A bipartite input will be denoted as an ordered pair $(first,$ $second)$, where $first$ and $second$ denote the first and second parts of the input. The two partitions where $first$ and $second$ are provided will be called the first and the second partition respectively.

A convention for dealing with the following issues for a Turing machine that takes a bipartite input is called a \textit{tape convention}: where to position the head initially, how the tape is divided into disjoint partitions, and where to place each part of a bipartite input in its partition. 

\begin{flushleft}
\textbf{Definition 3}. A tape convention refers to  a precise specification of the following:
\end{flushleft}
\begin{itemize}
\item[a)] a specific address as the initial head position, 
\item[b)] division of the tape squares into two disjoint partitions, and 
\item[c)] location of each part of a bipartite input in its corresponding partition
\end{itemize}

As an example, a tape convention  may, quite arbitrarily, specify that a) the tape head is to be initially positioned at the address 0, b) the first partition consists of those squares at addresses $<$ 0, and the second partition consists of those squares at addresses $\geq$ 0, and c) the first part of a bipartite input is to be located in the squares at the addresses $\cdots$  $-3,$ $-2,$ $-1$, with the rightmost symbol located at the address -1, and the second part of a bipartite input is to be located in the squares at the addresses $0,$ $1,$ $2,$ $\cdots$, with the leftmost symbol located at the address 0. 	
\begin{flushleft}
\textbf{Definition 4}. An execution of a Turing machine M on a  bipartite input $(first, second)$ is called a \textit{run} of M and is denoted by M$(first,$ $second)$. An identical tape convention is adopted for all runs of a given Turing machine.
\end{flushleft}

A run of a Turing machine solves an \textbf{instance} of the general problem that the Turing machine is intended to solve. For example, if a machine M solves the Boolean satisfiability problem, then a run of M decides whether a specific Boolean formula is satisfiable.

A tape convention for a Turing machine is analogous to an input convention assumed by a computer program: where the program's input is provided and how the input data are organized. That an identical tape convention is adopted for all runs of a given Turing machine is analogous to that an identical input convention is assumed by all executions of a given program. Since the same tape convention is adopted for all runs of a given Turing machine, each run executes with the same initial head position and with the tape divided into partitions in the same way. Besides, if two runs are given the same first (or second) part of a bipartite input, the initial content of the first (or second, respectively) partition for one run will be identical to that for the other run: that is, for every address $x$ in the first (or second, respectively) partition, the $x^{th}$ square for one run will initially bear the same symbol (possibly $blank$) as the $x^{th}$ square for the other run.

To illustrate bipartite inputs for multiple runs of a Turing machine, let us look at an example. Let M$(first_1,$ $second_1)$, M$(first_2,$ $second_2)$, M$(first_1,$ $second_2)$, M$(first_2,$ $second_1)$ be four runs of a machine M, where $first_1=$ $a_1a_2a_3$, $second_1=$ $b_1b_2b_3b_4$,  $first_2=$ $c_1c_2c_3c_4$, and $second_2=$ $d_1d_2d_3$, where each of $a_i$, $b_i$, $c_i$ and $d_i$ is a symbol in the tape alphabet of M. Suppose the adopted tape convention specifies, rather arbitrarily, that
\begin{itemize}
\item the first partition consists of those squares at addresses $< 15$ and the second partition consists of those at addresses $\geq 15$, and
\item the rightmost symbol of the string for the first part of each bipartite input is located at the address 14 and the leftmost symbol of the string for the second part is located at the address 15.
\end{itemize}

\noindent
The following table (Table 1) shows how the four bipartite inputs are provided for the four runs.

\begin{center}
\begin{tabular}{l|c|c|c|c|c|c|c|c|c|c|c|c|}
\cline{2-13}
& \multicolumn{6}{ |c| }{First partition}& \multicolumn{6}{ c|}{Second partition}\\ 
\hline
\multicolumn{1}{ |c| }{Addresses}&09&10&11&12&13&14&15&16&17&18&19&20 \\ 
\noalign{\hrule height 2pt}
\multicolumn{1}{ |c| }{M$(first_1,$ $second_1)$}& & & &$a_1$&$a_2$&$a_3$&$b_1$&$b_2$&$b_3$&$b_4$& &  \\ 
\hline
\multicolumn{1}{ |c| }{M$(first_2,$ $second_2)$}& & &$c_1$&$c_2$&$c_3$&$c_4$&$d_1$&$d_2$&$d_3$& & & \\ 
\hline
\multicolumn{1}{ |c| }{M$(first_1,$ $second_2)$}& & & &$a_1$&$a_2$&$a_3$&$d_1$&$d_2$&$d_3$&   & &  \\ 
\hline
\multicolumn{1}{ |c| }{M$(first_2,$ $second_1)$}& & &$c_1$&$c_2$&$c_3$&$c_4$&$b_1$&$b_2$&$b_3$&$b_4$ & & \\ 
\hline
\multicolumn{13}{ c }{Table 1. Bipartite Inputs for Multiple Runs} \\
\end{tabular}
\end{center}
\begin{flushleft}
Since the same tape convention is adopted for these runs, the first partition for M$(first_1,$ $second_1)$ is identical to that for M$(first_1,$ $second_2)$ because the two runs have the same first part $first_1$ in their bipartite inputs. Similarly, the second partition for M$(first_2,$ $second_2)$ is identical to that for M$(first_1,$ $second_2)$. So is the first partition for M$(first_2,$ $second_2)$ to that for M$(first_2,$ $second_1)$, and so too is the second partition for M$(first_1,$ $second_1)$ to that for M$(first_2,$ $second_1)$.
\end{flushleft} 
\begin{flushleft}
\textbf{Definition 5}. An \textit{execution path}, or simply a \textit{path}, is a sequence of 0 or more moves that a Turing machine may make as the machine executes its finite program, beginning from its start state.  A path is either \textit{terminated} or \textit{open}. If the start state is a halting state, then the null path is  terminated; otherwise, the null path is open. A non-null terminated path ends with a move whose destination state is a halting state, and an non-null open path ends with a move whose destination state is a non-halting state. A machine that serially makes the entire sequence of moves of a path is said to \textit{follow} the path. The first move that a machine makes after following an open path P is said to \textit{immediately succeed} the path P. A move that immediately succeeds a path P is called an \textit{immediate successor move} to P. 
\end{flushleft}
\begin{flushleft}
\textbf{Lemma 1}. For any Turing machine with $k$ symbols in its tape alphabet, there are no more than $k$ alternative immediate successor moves to any open path. 
\end{flushleft}
\begin{flushleft}
\textbf{Proof}. Let M be a Turing machine with $k$ symbols in its tape alphabet, P be an open path that M may follow, and $q$ be the non-halting state that M will be in at the end of following the path P. Since there are $k$ distinct tape symbols, there can be no more than $k$ distinct state-symbol pairs with $q$ as the state. Since the transition function of any Turing machine maps one or more distinct state-symbol pairs to a move, there are no more than $k$ distinct moves that may immediately succeed the path P. That is, the path P has no more than $k$ alternative immediate successor moves. \textbf{Q.E.D.}
\end{flushleft}
\begin{flushleft}
\textbf{Lemma 2}. For any Turing machine with $k$ symbols in its tape alphabet and for any integer $m \geq 0$, the sum of the following two numbers is no more than $k^m$.
\end{flushleft}
\begin{itemize}
\item[a)]the number of distinct open paths of $m$ moves and 
\item[b)]the number of distinct terminated paths of $m$ \textit{or fewer} moves.
\end{itemize} 
\begin{flushleft}
\textbf{Proof}. We will prove the lemma by induction. The null path, which can be either open or terminated, is the only one ($=k^0$) path of 0 move (basis of the induction). Suppose that there are $p$ distinct open paths of $i$ moves, there are $t$ distinct terminated paths of $i$ or fewer moves and  $p+t \leq k^i$  (inductive hypothesis). By Lemma 1, each of the $p$ open paths has no more than $k$ alternative immediate successor moves. Hence, by appending each of the $p$ open path  with each of its alternative immediate successor moves, we can form no more than $p k$ distinct paths of $i+1$ moves from the $p$ open paths and their alternative immediate successor moves. Of such paths of $i+1$ moves, some may remain open while the others become terminated. No new path can be formed from the $t$ terminated paths, which have no successor move. Hence the sum of the number of distinct open paths of $i+1$ moves and that of distinct terminated paths of $i+1$ or fewer moves is $p k+t$, which is no more than $k^{i+1}$ since by the inductive hypothesis $p+t \leq k^i$. \textbf{Q.E.D.}   
\end{flushleft}
\begin{flushleft}
\textbf{Lemma 3}. If two runs M$(first_1,$ $second_1)$ and M$(first_2,$ $second_2)$ of a Turing machine M follow a common terminated path P, then the run M$(first_1,$ $second_2)$ must follow the same terminated path P.
\end{flushleft}
\begin{flushleft}
\textbf{Proof}. We will prove the lemma by contradiction. Let the path P, which the two runs M$(first_1,$ $second_1)$ and M$(first_2,$ $second_2)$ follow, be the sequence of moves $P_1P_2$ $\cdots$ $P_p$, let the run M$(first_1,$ $second_2)$ follow the path Q, and let Q be the sequence of moves $Q_1Q_2$ $\cdots$ $Q_q$.  Assume that the path Q is different from the path P. We will derive a contradiction to this assumption. Since Q is different from P, there exists an integer $i$  such that the move $Q_i$ is different from the move $P_i$. Of such integers, there must be a least one. Let $m$ be the least such integer. Since $m$ is the smallest integer such that the move $Q_m$ is different from the move $P_m$, the sequence $P_1 $ $\cdots$ $ P_{m-1}$ is identical to the sequence $Q_1$ $\cdots$ $ Q_{m-1}$. Let T be the point in P and in Q between their $(m-1)^{st}$ move and their $m^{th}$ move. Let the three runs proceed to the point T, where each run has completed the common sequence of moves $P_1 $ $\cdots$ $ P_{m-1}$ but has not made the $m^{th}$ move, which is the move $P_m$ for M$(first_1,$ $second_1)$ and M$(first_2,$ $second_2)$ or the move $Q_m$ for M$(first_1,$ $second_2)$. At point T, all three runs are in some common internal state $q$. This is because if the common path $P_1 $ $\cdots$ $ P_{m-1}$ is null, then the common state $q$ is the start state; otherwise, the common state $q$ is the destination state of the move $P_{m-1}$, the last move of the path $P_1 $ $\cdots$ $ P_{m-1}$. Besides, at the point T, all three runs have a common head position. This is because all three runs begin execution with the same initial head position and then make an identical sequence of head movements as they follow the common path $P_1 $ $\cdots$ $ P_{m-1}$ to the point T. Let $x$ be the address of this common head position for all three runs at the point T. Consider the $x^{th}$ tape square for the run M$(first_1,$ $second_2)$: the square either has been written by a move in the common path $P_1 $ $\cdots$ $ P_{m-1}$, or it has not. In either case, $Q_m$ and $P_m$ can be shown to be the same move, as detailed below.
\end{flushleft}
\begin{itemize}
\item[A)] Suppose the $x^{th}$ tape square for M$(first_1,$ $second_2)$ has been written by a move in the path $P_1 $ $\cdots$ $ P_{m-1}$. Since, in following the common path $P_1 $ $\cdots$ $ P_{m-1}$ to the point T, all three runs make an identical sequence of head movements and perform an identical sequence of write operations, at the point T the $x^{th}$ tape square for each of the three runs must have been written with the same symbol. Hence, at the point T, all three runs scan the same symbol. Since all three runs are also in the same internal state at the point T, they have the same state-symbol pair, which the transition function maps to the same $m^{th}$  move for  all three runs. That is, the move $Q_m$ is the same as the move $P_m$.
\item[B)] Suppose the $x^{th}$ tape square for M$(first_1,$ $second_2)$ has \textbf{not} been written by a move along the common path $P_1 $ $\cdots$ $ P_{m-1}$. The address $x$ is either in the first partition or in the second. In either case, $Q_m$ and $P_m$ can be shown to be the same move, as detailed below. 
\begin{itemize}
\item[B.1)]Suppose $x$ is in the first partition. Since M$(first_1,$ $second_2)$  and M$(first_1,$ $second_1)$  have an identical first part in their bipartite inputs, both runs are given identical initial content in their first partition. Since the $x^{th}$ square has not been written along the path $P_1 $ $\cdots$ $ P_{m-1}$, at the point T the $x^{th}$ square for M$(first_1,$ $second_2)$ and the corresponding square for M$(first_1,$ $second_1)$ both bear the initial symbol (possibly \textit{blank}) at the address $x$ in the first partition. Hence, at the point T, M$(first_1,$ $second_2)$  and M$(first_1,$ $second_1)$ scan the same symbol. Since the two runs are also in a common internal state at the point T, they have the same state-symbol pair, which the transition function maps to the same $m^{th}$ move for M$(first_1,$ $second_2)$ and M$(first_1,$ $second_1)$. That is, the move $Q_m$ is the same as the move $P_m$.
	
\item[B.2)]Suppose $x$ is in the second partition. Similarly to case B.1, the  two runs M$(first_1,$ $second_2)$ and M$(first_2,$ $second_2)$ can be shown to have the same state-symbol pair at the pint T, which the transition function maps to the same $m^{th}$ move for M$(first_1,$ $second_2)$ and M$(first_2,$ $second_2)$. That is, the move $Q_m$ is the same as the move $P_m$.
\end{itemize}
\end{itemize}
\begin{flushleft}
Thus, there does not exist an integer $m$ such that $Q_m$ is different from $P_m$. In other words, the path P and the path Q are identical. \textbf{Q.E.D.}
\end{flushleft}

It is interesting to note that Lemma 3 holds no matter which address the head is initially positioned at, no matter how the tape is divided into disjoint partitions, and no matter where each part of a bipartite input is placed in its corresponding partition. In short, the lemma holds no matter what tape convention is adopted, as long as an identical tape convention is adopted for the runs involved.
\section{A Lower Bound for Satisfiability}

We now establish a lower bound on the number of moves required to decide Boolean satisfiability on a Turing machine. 
\begin{flushleft}
\textbf{Definition 6}. Let $x_1,$ $x_2$ $\cdots$ $x_n$ be the  variables of which the members of the set $B^n$ $\rightarrow$ $B$ are functions. A Boolean formula that  \textit{represents} or \textit{defines} a Boolean function $f:B^n$ $\rightarrow$ $B$ is a formula $\phi_f$ of the variables $x_1,$ $x_2$ $\cdots$ $x_n$ such that, for every assignment to the variables $x_1,$ $x_2$ $\cdots$ $x_n$, the formula $\phi_f$ evaluates to the same value as what the function $f$ evaluates to.
\end{flushleft}

A Boolean formula that represents a function $f$ will be denoted by $\phi_f$ here. However, the symbol $\phi$ without a subscript, or with a numerical subscript, such as $\phi_3$, will denote a Boolean formula without indicating the specific function that it represents.

\begin{flushleft}
\textbf{Definition 7}. Let $x_1,$ $x_2$ $\cdots$ $x_n$ be the  variables of which the members of the set $B^n$ $\rightarrow$ $B$ are functions. A \textit{full representation} of the set $B^n$ $\rightarrow$ $B$ is a set $E$ of Boolean formulas of the variables $x_1,$ $x_2$ $\cdots$ $x_n$ such that  every function $f:B^n$ $\rightarrow$ $B$ is represented by a formula $\phi_f$ $\in$ $E$. The set $E$ is said to \textit{fully represent} the set $B^n$ $\rightarrow$ $B$.
\end{flushleft}

\begin{flushleft}
\textbf{Definition 8}. The \textit{logical negation} of a function $g:B^n$ $\rightarrow$ $B$ is a function $\overline{g}:B^n$ $\rightarrow$ $B$ such that, for every assignment to the variables $x_1,$ $x_2$ $\cdots$ $x_n$, 
\end{flushleft}
\begin{center}
$\overline{g}(x_1,$ $x_2$ $\cdots$ $x_n)$ ==$\overline{g(x_1, x_2 \cdots x_n)}$.
\end{center}

The logical negation of a function $g$ is denoted by $\overline{g}$. For any function $g:B^n$ $\rightarrow$ $B$ and for every assignment, $g$ and $\overline{g}$ must evaluate to different values: one of them must evaluate to $false$ and the other to $true$. 
\begin{flushleft}
\textbf{Definition 9}. A run M$(\phi_1,$ $\phi_2)$ is said to \textit{decide the satisfiability of $\phi_1$ $\wedge$ $\phi_2$}, or to \textit{decide whether $\phi_1$ $\wedge$ $\phi_2$ is satisfiable}, if and only if the run accepts its input (i.e., halts in the accept state) if $\phi_1$ $\wedge$ $\phi_2$ is satisfiable and rejects its input (i.e., halts in the reject state) otherwise. A run M$(\phi_1,$ $\phi_2)$ is said to \textit{decide the falsifiability of $\phi_1$ $\vee$ $\phi_2$}, or to  \textit{decide whether $\phi_1$ $\vee$ $\phi_2$ is falsifiable}, if and only if the run accepts its input if $\phi_1$ $\vee$ $\phi_2$ is falsifiable and rejects its input otherwise.
\end{flushleft}

\begin{flushleft}
\textbf{Theorem 1}. Let $E$ be a full representation of the set  $B^n$ $\rightarrow$ $B$. There \textbf{does not exist} a Turing machine M with $k$ symbols in its tape alphabet such that, for every pair of formulas $\phi_1,$ $\phi_2$ $\in$ $E$, M$(\phi_1,$ $\phi_2)$ correctly decides  whether $\phi_1$ $\wedge$ $\phi_2$ is satisfiable by making fewer than $2^nlog_k2$ moves.
\end{flushleft} 
\textbf{Proof}. We will prove the theorem by contradiction. We first assume that there exists a Turing machine M with $k$ symbols in its tape alphabet such that, for every pair of formulas $\phi_1,$ $\phi_2$ $\in$ $E$, M$(\phi_1,$ $\phi_2)$ correctly decides the satisfiability of $\phi_1$ $\wedge$ $\phi_2$ by making fewer than $2^nlog_k2$ moves. In other words, M$(\phi_1,$ $\phi_2)$ will accept the input if $\phi_1$ $\wedge$ $\phi_2$ is satisfiable and reject the input otherwise, and M$(\phi_1,$ $\phi_2)$ will do so by making fewer than $2^nlog_k2$  moves. The rest of this proof will derive a contradiction to this assumption.

Since $E$ fully represents the set $B^n$ $\rightarrow$ $B$, every function $f:B^n$ $\rightarrow$ $B$ and its logical negation $\overline{f}:B^n$ $\rightarrow$ $B$ are represented by some Boolean formulas $\phi_f,$ $\phi_{\overline{f}}$ $\in$ $E$. Let S be a set containing, for each distinct function $f:B^n$ $\rightarrow$ $B$, one run of M with $(\phi_f,$ $\phi_{\overline{f}})$ as its bipartite input.  In other words, for each function $f:B^n$ $\rightarrow$ $B$, S contains the run M$(\phi_f,$ $\phi_{\overline{f}})$, which is to decide whether the formula  $\phi_f$ $\wedge$ $\phi_{\overline{f}}$ is satisfiable. Since there are $F=$ $2^{2^n}$ distinct functions in the set $B^n$ $\rightarrow$ $B$, the set S has $F$ runs of the machine M. 

The set S may seem expensive to implement in terms of computing resources. However, S will only be used to prove that logically the Turing machine M does not exist. An actual implementation of S is not needed. 

Since, for every function $f:B^n$ $\rightarrow$ $B$ and for every assignment, either the function $f$ or its logical negation $\overline{f}$ evaluates to $false$, and since $\phi_f,$ $\phi_{\overline{f}}$ $\in$ $E$ represent $f$ and $\overline{f}$, for every assignment either $\phi_f$ or $\phi_{\overline{f}}$ evaluates $false$. Therefore, the formula $\phi_f$ $\wedge$ $\phi_{\overline{f}}$ is $false$ for every assignment and, thus,	 is not satisfiable. Hence, every run in the set S must eventually reject its input. By our assumption on M, every run in S must follow a terminated path of $2^nlog_k2$ $-1$ or fewer moves and reject its input. 

By Lemma 2, there are no more than $k^m$ terminated paths of $m$ or fewer moves. Since each run in S follows a terminated path of $2^nlog_k2$ $-1$ or fewer moves, by Lemma 2 there are no more than the following number of terminated paths that the runs in S may follow.
\begin{center}
$k^{(2^nlog_k2)-1} = k^{(2^nlog_k2)} k^{-1} = k^{(log_k2)2^n}  k^{-1} = (k^{log_k2})^{2^n} k^{-1} = 2^{2^n} k^{-1} = F k^{-1}  = F/k$
\end{center}

To summarize, each of the $F =$ $2^{2^n}$ runs in the set S follows a terminated path of $2^nlog_k2$ $-1$ or fewer moves to reject its input, but there are no more than $F/k$ such paths. Therefore, there is at least one such path that multiple runs in S follow. Let P be a path that multiple runs in S follow and let M$(\phi_g,$ $\phi_{\overline{g}})$  and M$(\phi_h,$ $\phi_{\overline{h}})$ be two runs in S that follow the path P. Since S contains one run of M for each distinct Boolean function of $n$ variables, $g$ and $h$ must be different functions.  Since, as discussed previously, all runs in S must reject their inputs,  both M$(\phi_g,$ $\phi_{\overline{g}})$  and M$(\phi_h,$ $\phi_{\overline{h}})$ must reject their inputs. By Lemma 3,  two other runs, M$(\phi_g,$ $\phi_{\overline{h}})$  and M$(\phi_h,$ $\phi_{\overline{g}})$, which are not in S, must follow the same path P and \textbf{reject} their inputs, as the two runs M$(\phi_g,$ $\phi_{\overline{g}})$  and M$(\phi_h,$ $\phi_{\overline{h}})$ do. 

Now let us derive a contradiction to the assumption that the Turing machine M exists. Since $g$ and $h$ are different Boolean functions, there exists an assignment $s$ that makes $g$ and $h$ evaluate to different values. Hence, the assignment s will make $g$ and $\overline{h}$ evaluate to the same value. If both $g$ and $\overline{h}$ evaluate to $true$ on the assignment $s$, so will both $\phi_g$ and $\phi_{\overline{h}}$, since $\phi_g,$ and $\phi_{\overline{h}}$ represent $g$ and $\overline{h}$. Thus, $\phi_g$ $\wedge$ $\phi_{\overline{h}}$ is satisfiable. On the other hand, if $g$ and $\overline{h}$ evaluate to $false$ on the assignment $s$, then $h$ and $\overline{g}$ will evaluate to $true$ on the assignment $s$ and so will $\phi_h$ and $\phi_{\overline{g}}$, since $\phi_h$ and $\phi_{\overline{g}}$  represent $h$ and $\overline{g}$. Thus, $\phi_h$ $\wedge$ $\phi_{\overline{g}}$ is satisfiable. Therefore, at least one of the two formulas $\phi_g$ $\wedge$ $\phi_{\overline{h}}$  and $\phi_h$ $\wedge$ $\phi_{\overline{g}}$ is satisfiable and, thereby, at least one of the two runs M$(\phi_g,$ $\phi_{\overline{h}})$  and M$(\phi_h,$ $\phi_{\overline{g}})$ \textbf{should accept} its input. However, as discussed previously, by Lemma 3 both M$(\phi_g,$ $\phi_{\overline{h}})$  and M$(\phi_h,$ $\phi_{\overline{g}})$ \textbf{reject} their inputs. That is, by Lemma 3, at least one of the two runs M$(\phi_g,$ $\phi_{\overline{h}})$  and M$(\phi_h,$ $\phi_{\overline{g}})$ \textbf{incorrectly} rejects its input. This contradicts our assumption that the machine M exists such that, for every pair of formulas $\phi_1,$ $\phi_2$ $\in$ $E$, M$(\phi_1,$ $\phi_2)$ \textbf{correctly} decides the satisfiability of $\phi_1$ $\wedge$ $\phi_2$ by making fewer than $2^nlog_k2$ moves. \textbf{Q.E.D.}

\null

By Theorem 1, for any Turing machine M with $k$ symbols in its tape alphabet, there is at least one pair of formulas $\phi_1$ and $\phi_2$ in any full representation of $B^n$ $\rightarrow$ $B$ such that M$(\phi_1,$ $\phi_2)$ cannot correctly decide the satisfiability of $\phi_1$ $\wedge$ $\phi_2$  by making fewer than $2^nlog_k2$ moves. In other words, $2^nlog_k2$ is a lower bound on the number of moves needed.

Like Lemma 3, Theorem 1 holds regardless of the initial head position, the way the tape is divided into disjoint partitions, and the location of each part of a bipartite input in its corresponding partition. In short, the theorem holds no matter what tape convention is adopted, as long as an identical tape convention is adopted for the runs involved. Besides, it should be noted that the proof for Theorem 1 does not rely on a specific representation of Boolean functions. As a result, the lower bound applies to the problem of deciding whether the conjunction of a pair of $n$-variable Boolean functions has a satisfying assignment, even if the two conjuncts are represented in the input as some expressions other than the form of Boolean formulas introduced in Section 2.1. 

Since there are many different Boolean formulas that represent a given Boolean function, there are many full representations of the set $B^n$ $\rightarrow$ $B$. It is interesting to note that Theorem 1  holds for any full representation $E$, even if $E$ consists solely of minimized Boolean formulas that are derived by a Boolean minimization method.

Since the set $B^n$ $\rightarrow$ $B$ can be fully represented by a set of CNF formulas, the lower bound holds even if the conjuncts $\phi_1$ and $\phi_2$ are limited to CNF formulas. 
\begin{flushleft}
\textbf{Corollary 1.1}. Let $E$ be a set of CNF formulas that fully represents  $B^n$ $\rightarrow$ $B$. There \textbf{does not exist} a Turing machine M with $k$ symbols in its tape alphabet such that, for every pair of formulas $\phi_1,$ $\phi_2$ $\in$ $E$, M$(\phi_1,$ $\phi_2)$ correctly decides  whether the CNF formula $\phi_1$ $\wedge$ $\phi_2$ is satisfiable by making fewer than $2^nlog_k2$ moves.
\end{flushleft}

By duality, Corollary 1.2 follows from Theorem 1:
\begin{flushleft}
\textbf{Corollary 1.2}. Let $E$ be a full representation of $B^n$ $\rightarrow$ $B$. There \textbf{does not exist} a Turing machine M with $k$ symbols in its tape alphabet such that, for every pair of formulas $\phi_1,$ $\phi_2$ $\in$ $E$, M$(\phi_1,$ $\phi_2)$ correctly decides  whether $\phi_1$ $\vee$ $\phi_2$ is falsifiable by making fewer than $2^nlog_k2$ moves.
\end{flushleft}

By duality, Corollary 1.3 follows from Corollary 1.1.
\begin{flushleft}
\textbf{Corollary 1.3}. Let $E$ be a set of DNF formulas that fully represents $B^n$ $\rightarrow$ $B$. There \textbf{does not exist} a Turing machine M with $k$ symbols in its tape alphabet such that, for every pair of formulas $\phi_1,$ $\phi_2$ $\in$ $E$, M$(\phi_1,$ $\phi_2)$ correctly decides  whether the DNF formula $\phi_1$ $\vee$ $\phi_2$ is falsifiable by making fewer than $2^nlog_k2$ moves. 
\end{flushleft}
\section{Restricted Formulas}

Theorem 1 requires that the two conjuncts $\phi_1$ and $\phi_2$ be members of a full representation of $B^n$ $\rightarrow$ $B$. Since the following widely known sets of restricted formulas of $n$ variables do not fully represent $B^n$ $\rightarrow$ $B$,  Theorems 1 does not apply if the two conjuncts are limited to $n$-variable formulas from any of these sets: XOR-SAT, HORN-SAT, 2CNF, and 3CNF. Polynomial-time algorithms to decide 2CNF satisfiability,  XOR-SAT, and HORN-SAT are known. The next theorem establishes a lower bound of $2^nlog_k2$ moves for 3CNF satisfiability.

\begin{flushleft}
\textbf{Definition 10}. Let $E_1$ and $E_2$ be sets of Boolean formulas. A \textit{satisfiability-preserving} mapping from $E_1$ to $E_2$ is a function $t:E_1$ $\rightarrow$ $E_2$ such that, for every formula $\phi$ $\in$ $E_1$, the image $t(\phi)$ $\in$ $E_2$ is satisfiable if and only if $\phi$ is satisfiable. The function $t$ is said to \textit{preserve satisfiability}.
\end{flushleft}
\begin{flushleft}

\textbf{Definition 11}. Let $E_1$ and $E_2$ be sets of Boolean formulas. A mapping from $E_1$ to $E_2$ that \textit{preserves satisfiability over conjunction} is a function  $t:E_1$ $\rightarrow$ $E_2$ such that, for every pair of formulas $\phi_1,$ $\phi_2$ $\in$ $E_1$, the formula $t(\phi_1)$ $\wedge$ $t(\phi_2)$ is satisfiable if and only if $\phi_1$ $\wedge$ $\phi_2$ is satisfiable. The function $t$ is said to be \textit{satisfiability-preserving over conjunction}.
\end{flushleft}

\begin{flushleft}
\textbf{Definition 12}. A set $E$ of Boolean formulas is said to be a \textit{satisfiability representation} of the set $B^n$ $\rightarrow$ $B$ if and only if there exist a full representation $E_1$ of the set $B^n$ $\rightarrow$ $B$ and a function $t:E_1$ $\rightarrow$ $E$ that preserves satisfiability over conjunction. The set $E$ is  said to \textit{satisfiability-represent} the set $B^n$ $\rightarrow$ $B$.
\end{flushleft}

\begin{flushleft}
\textbf{Theorem 2}. Let $E$ be a satisfiability representation of $B^n$ $\rightarrow$ $B$. There \textbf{does not exist} a Turing machine M with $k$ symbols in its tape alphabet such that, for every pair of formulas $\phi_1,$ $\phi_2$ $\in$ $E$, M$(\phi_1,$ $\phi_2)$ correctly decides  whether the formula $\phi_1$ $\wedge$ $\phi_2$ is satisfiable by making fewer than $2^nlog_k2$ moves.
\end{flushleft}

\begin{flushleft}
\textbf{Proof}. Our proof for Theorem 2 is essentially identical to that for Theorem 1, with the following adaptions:
\end{flushleft}
\begin{itemize} [align=left,leftmargin=*]
\item[1.] The proof assumes that  there exists a Turing machine M such that, for every pair of formulas $\phi_1,$ $\phi_2$ $\in$ $E$, M$(\phi_1,$ $\phi_2)$ correctly decides  the satisfiability of  $\phi_1$ $\wedge$ $\phi_2$ by making fewer than $2^nlog_k2$ moves. 

\item[2.] Since $E$ satisfiability-represents $B^n$ $\rightarrow$ $B$, there is a set $E_1$ that is a full representation of $B^n$ $\rightarrow$ $B$ and there is a function  $t:E_1$ $\rightarrow$ $E$ that preserves satisfiability over conjunction. Let the set S contain, for each function $f:B^n$ $\rightarrow$ $B$, one run of M with $(t(\phi_f),$ $t(\phi_{\overline{f}}))$ as its bipartite input, where $\phi_f,$ $\phi_{\overline{f}}$ $\in$ $E_1$ and, hence, $t(\phi_f),$ $t(\phi_{\overline{f}})$ $\in$ $E$.

\item[3.] For every function $f:B^n$ $\rightarrow$ $B$ and for every assignment, one of $f$ and $\overline{f}$ evaluates to $false$. Since $\phi_f$ and $\phi_{\overline{f}}$ represent $f$ and $\overline{f}$,  for every assignment one of $\phi_f$ and $\phi_{\overline{f}}$ evaluates to $false$. Hence, $\phi_f$ $\wedge$ $\phi_{\overline{f}}$ is false for all assignments and, thus, is not satisfiable. Since $t$ is satisfiability-preserving over conjunction,  the formula  $t(\phi_f)$ $\wedge$ $t(\phi_{\overline{f}})$ is not satisfiable. Hence, every run in S must eventually reject its input.

\item[4.] To derive a contradiction, let M$(t(\phi_g),$ $t(\phi_{\overline{g}}))$ and  M$(t(\phi_h),$ $t(\phi_{\overline{h}}))$ be two runs in S that follow a common path P of fewer than $2^nlog_k2$ moves to reject their inputs - as deptailed in the proof for Theorem 1, there must be at least two such runs in S. By Lemma 3,  the two runs M$(t(\phi_g),$ $t(\phi_{\overline{h}}))$ and M$(t(\phi_h),$ $t(\phi_{\overline{g}}))$, which are not in S, must follow the same execution path P to  reject their inputs, as the two runs   M$(t(\phi_g),$ $t(\phi_{\overline{g}}))$ and  M$(t(\phi_h),$ $t(\phi_{\overline{h}}))$ do. Since $g$ and $h$ are different Boolean functions, there exists an assignment $s$ that makes $g$ and $h$ evaluate to different values. Therefore, $g$ and $\overline{h}$ evaluate to the same value on the assignment $s$. If both $g$ and $\overline{h}$ evaluate to $true$ on the assignment $s$, then so will both $\phi_g$ and $\phi_{\overline{h}}$ since $\phi_g$ and $\phi_{\overline{h}}$ represent $g$ and $\overline{h}$. Hence, $\phi_g$ $\wedge$ $\phi_{\overline{h}}$ is satisfiable.  Since $t$ is satisfiability-preserving over conjunction, $t(\phi_g)$ $\wedge$ $t(\phi_{\overline{h}})$ is satisfiable too. On the other hand, if both $g$ and $\overline{h}$  evaluate to $false$ on the assignment $s$, then both $h$ and $\overline{g}$ evaluate to $true$ on the assignment $s$, and $t(\phi_h)$ $\wedge$ $t(\phi_{\overline{g}})$ can be similarly shown to be satisfiable. So, at least one of the formulas $t(\phi_g)$ $\wedge$ $t(\phi_{\overline{h}})$ and $t(\phi_h)$ $\wedge$ $t(\phi_{\overline{g}}))$ is satisfiable. That is, at least one of the two runs M$(t(\phi_g),$ $t(\phi_{\overline{h}}))$ and M$(t(\phi_h),$ $t(\phi_{\overline{g}}))$ \textbf{should accept} its input. However, as discussed previously, by Lemma 3 both M$(t(\phi_g),$ $t(\phi_{\overline{h}}))$ and M$(t(\phi_h),$ $t(\phi_{\overline{g}}))$ \textbf{reject} their inputs. That is, by Lemma 3, at least one  of the two runs M$(t(\phi_g),$ $t(\phi_{\overline{h}}))$ and M$(t(\phi_h),$ $t(\phi_{\overline{g}}))$ \textbf{incorrectly} rejects its input. This contradicts the assumption stated above in item 1. \textbf{Q.E.D.}
\end{itemize}

We give an example of a set of restricted Boolean formulas that satisfiability-represents $B^n$ $\rightarrow$ $B$.  It is well known that the problem of CNF satisfiability can be reduced to 3CNF satisfiability, e.g., [2,3]. Specifically, when this reduction is applied to a CNF formula $C_1$ $\wedge$ $C_2$, where   $C_1$ and $C_2$ are CNF formulas, the  reduction yields a formula $t(C_1)$ $\wedge$ $t(C_2)$ as the resultant 3CNF formula, where $t(C_1)$ and $t(C_2)$ are 3CNF formulas and are derived by applying the reduction to $C_1$ and $C_2$ respectively. The formulas $t(C_1)$ and $t(C_2)$ are satisfiable if and only if $C_1$ and $C_2$ are, respectively, and the resultant 3CNF formula $t(C_1)$ $\wedge$ $t(C_2)$ is satisfiable if and only if the original CNF formula $C_1$ $\wedge$ $C_2$ is. This reduction  introduces distinct new variables into the resultant 3CNF formulas. With the new variables being distinct, this reduction defines a mapping from CNF formulas to 3CNF formulas that is satisfiability-preserving over conjunction. Let $E_1$ be a set of CNF formulas that fully represents $B^n$ $\rightarrow$ $B$. This reduction can be used to transform each CNF formula in $E_1$ into a 3CNF formula. Let $E$ be the set of the resultant 3CNF formulas. The set $E$ satisfiability-represents the set $B^n$ $\rightarrow$ $B$.

The following corollary directly follows from Theorem 2.

\begin{flushleft}
\textbf{Corollary 2.1}. Let $E$ be a set of 3CNF formulas that satisfiability-represents $B^n$ $\rightarrow$ $B$. There \textbf{does not exist} a Turing machine M with $k$ symbols in its tape alphabet such that, for every pair of 3CNF formulas $\phi_1,$ $\phi_2$ $\in$ $E$, M$(\phi_1,$ $\phi_2)$ correctly decides  whether the 3CNF formula $\phi_1$ $\wedge$ $\phi_2$ is satisfiable by making fewer than $2^nlog_k2$ moves. 
\end{flushleft}

Similarly, there is a reduction from the problem of DNF falsifiability to 3DNF falsifiability [1].  By duality, the following corollary follows from Corollary 2.1.  The term \textit{falsifiability-represent} is the dual of the term satisfiability-represent defined previously. A detailed definition of the term falsifiability-represent parallels Definitions 11-12.

\begin{flushleft}
\textbf{Corollary 2.2}. Let $E$ be a set of 3DNF formulas that falsifiability-represents $B^n$ $\rightarrow$ $B$. There \textbf{does not exist} a Turing machine M with $k$ symbols in its tape alphabet such that, for every pair of 3DNF formulas $\phi_1,$ $\phi_2$ $\in$ $E$, M$(\phi_1,$ $\phi_2)$ correctly decides  whether the 3DNF formula $\phi_1$ $\vee$ $\phi_2$ is falsifiable by making fewer than $2^nlog_k2$ moves.
\end{flushleft}

\section{Discussion}

It is interesting to note that, for a Turing machine using a binary tape alphabet, the lower bound becomes $2^n$ because when $k=2$, $2^nlog_k2=$ $2^nlog_22=$ $2^n$. When a binary alphabet is used, a tape square can be in one of only two possible states, for example, a tape square can be either $blank$ or $nonblank$. Just as binary strings can be used to represent various kinds of information, permutations of $blank$ and $nonblank$ squares can be so used too. 

More generally, a tape alphabet may consist of a constant number of tape symbols. For the Boolean satisfiability problem, a tape alphabet may, for example, consist of symbols to denote Boolean values, Boolean operators, parentheses, and the $blank$ symbol, as well as symbols to form strings to represent identifiers. With a constant number of distinct symbols in the tape alphabet,  it is still possible to represent an unlimited number of identifiers, values, and formulas as strings formed from the alphabet. With the number of distinct symbols in the tape alphabet being a constant,  the lower bound is $c2^n$ moves, where $c$ is the constant $log_k2$.

Since the lower bound established in this paper does not depend on sequentiality of access to the tape squares, the lower bound will hold even if a Turing machine is capable of non-sequential access to the tape squares. For example, even if a move is allowed to specify that the read/write head move to a square at a certain  address, or that the read/write head skip a certain number of squares to the left or to the right, the lower bound of $2^nlog_k2$ moves will still hold. The proofs only require straightforward adaptions to accommodate this flexibility.

Additionally, the lower bound will still hold if a tape convention divides the tape into more than two disjoint partitions, although only two partitions are used to hold a bipartite input. The proofs only require minor adaptions to accommodate this flexibility.
\begin{flushleft}
\textbf{References}
\end{flushleft}
\begin{itemize}[align=left,leftmargin=*]
\item[1.] Cook, S.A. The complexity of theorem proving procedures. In \textit{Proceedings, Third Annual ACM Symposium on the Theory of Computing} (1971), pp. 151-158.
\item[2.] Hopcroft, J.E., Ullman, J.D., \textit{Introduction to Automata Theory, Languages, and Computation}. Addison-Wesley, 1979.
\item[3.] Sipser, M. \textit{Introduction to the Theory of Computation}. 2nd ed. Thomson Course Technology, 2006.
\item[4.] Turing, A.M. On Computable Numbers, with an Application to the Entscheidungs problem. In \textit{Proceedings of the London Mathematical Society} (1936), pp.230-265.
\end{itemize}
\end{document}